\documentclass[11pt]{article}

\usepackage{amsmath,amssymb}
\usepackage{slashed}
\usepackage{xcolor} 
\usepackage{graphicx}
\usepackage{dcolumn} 
\usepackage{bm} 
\usepackage{epsfig}
\usepackage{epstopdf}
\usepackage{grffile}
\usepackage{color}
\usepackage{colordvi}
\usepackage{amsmath,amssymb}
\usepackage{rotating}
\usepackage{lscape}
\usepackage{cite}
\usepackage{float}
\usepackage{hyperref}

\def\Re{{\cal R \mskip-4mu \lower.1ex \hbox{\it e}\,}}
\def\Im{{\cal I \mskip-5mu \lower.1ex \hbox{\it m}\,}}

\def\tev{\,{\ifmmode\mathrm {TeV}\else TeV\fi}}
\def\gev{\,{\ifmmode\mathrm {GeV}\else GeV\fi}}
\def\mev{\,{\ifmmode\mathrm {MeV}\else MeV\fi}}

\begin{document}
\begin{center}
\vspace*{15mm}
\vspace{1cm}
{\Large \bf Probing Higgs boson couplings in H+$\gamma$ production at the LHC}
\vspace{1cm}

{\bf Hamzeh Khanpour$^{\S,\dagger}$,  Sara Khatibi$^{\dagger}$,   and Mojtaba Mohammadi Najafabadi$^{\dagger}$ }

 \vspace*{0.5cm}

{\small\sl 
$^{\S}$Department of Physics, University of Science and Technology of Mazandaran, P.O.Box 48518-78195, Behshahr, Iran }\\
$^{\dagger}$School of Particles and Accelerators, Institute for Research in Fundamental Sciences (IPM) P.O. Box 19395-5531, Tehran, Iran  \\

\vspace*{.2cm}
\end{center}

\vspace*{10mm}

%
%
\begin{abstract}\label{abstract}
In this paper, we examine the potential of  Higgs boson production associated with a photon at the LHC
to probe the new physics effects in the framework of the standard model effective field theory.
It is shown that the differential kinematic distributions such as 
photon transverse momentum and invariant mass of Higgs+$\gamma$ in Higgs 
associated production are powerful variables to explore the coefficients of dimension six operators.
The analysis is performed in the decay channel of Higgs boson into a $b\bar{b}$ pair
including the main sources of background processes and a realistic simulation of the detector effects.
We provide constraints at $95\%$ confidence level on the Wilson coefficients of dimension-six operators affecting Higgs
boson plus a photon production. We show to what extent these limits could be improved 
at the high luminosity LHC.
The effect of these constraints on a well-motivated beyond standard model scenario is presented.
\end{abstract}

\newpage

\section{Introduction}\label{sec:intro}

After the discovery of Higgs boson by the LHC experiments \cite{atlash,cmsh}, standard model (SM) 
has been found to be a successful theory to describe the interactions among the fundamental particles
at the electroweak scale. However, it has some shortcomings which leads us to construct new models beyond the SM.
Various models such as supersymmetric SM, extra-dimensions, two-Higgs doublet models etc. have been 
suggested to solve the SM problems.  For instance, supersymmetric extension of SM provides the possibility to stabilize the Higgs boson mass
from the ultra-violet divergencies through the contributions of the supersymmetric partners of the SM particles \cite{susy}. 
Most of these theories lead to some modifications of the SM parameters. In particular, the Higgs boson couplings 
are affected by several extensions of the SM. Nevertheless,  all the experimental  measurements done by the LHC 
experiments are in agreement with the SM predictions  and so far no significant deviation with respect
to the SM expectations have been found \cite{r1,r2,r220,r221}.
As a result, any new degrees of freedom are expected to be well separated  from the SM particles in mass \cite{r3,r4}.
Due to the presence of several beyond SM scenarios and in some cases similar 
experimental signatures, a useful way to search for new physics  could be done 
in a model independent way. In other words, the new physics effects could show up 
in the effective field theory extension of the SM which is composed of
an infinite series of higher-dimensional effective operators \cite{r5,r6,r7,r8,r9,r10}.
It is built based on the SM degrees of
freedom and its symmetries and it could be written by
adding new higher  than four dimension operators to the Lagrangian of the SM.
The leading contribution of the SM effective Lagrangian comes from
the operators of dimension-six that is based on a complete and non-redundant operator basis \cite{r11,r12, r13}.

From the phenomenological point of view, these operators 
can affect not only the signal strengths but also the differential distributions and angular observables as they
contain new vertex structures.
There are many effective operators that are
contributing to Higgs couplings which motivate us to look at 
 all possible Higgs involved processes at the LHC.
Studying all processes in which Higgs boson is involved 
such as Higgs+jets production, Higgs associated production, and
processes where Higgs is off-shell is necessary to provide information of 
all related new couplings. 
So far, there have been many studies for exploring new physics effects in
the Higgs boson sector at the particle colliders in the context of SM effective field theory 
\cite{r14,r15,r16,r17,r18,r19,r20,r21,r22,r23,r24,r25,r26,r27,r28,r29,r30,r31,r32,r33,r330,r331, r88, r89,r999, r1000, r1001, r1002, r1003, r1004,r1005,r1006, r1007,r1008,r1009}.

At the LHC, the Higgs production mostly proceeds through gluon-gluon fusion,
vector boson fusion (H+2 jets) and associated production with a $Z$ or a $W$ boson.
Higgs boson could also be produced associated with a pair of top or a single top quark with production rates 
substantially less than the  gluon-gluon fusion, vector boson fusion and associated production \cite{r34}. 
In addition to these processes, Higgs boson can be produced in association with a photon which has 
a tiny rate. However, studying this process is interesting as the photon in the final state
is a hard photon and is a handle to suppress the background processes. Also, 
models beyond the SM can affect the production cross section of this process \cite{r340}. 

In this article,  the power of associated H+$\gamma$ production and its differential 
distributions are used to constrain CP-conserving dimension-six operators.
Particularly, the photon transverse momentum and the  invariant mass of H+$\gamma$ system
are utilized to probe the new couplings effects. These observables receive significant contributions
from the dimension-six operators at large values, as these effective operators lead to high momentum 
transfers in associated Higgs production. We will show this channel would provide useful 
information on Higgs boson couplings and is worth to be looked for by the LHC experiments. 

In  section \ref{sec:framework}, we introduce the  dimension-six operators which affect
Higgs boson sector.  Section \ref{sec:smm} is dedicated to discuss the production of Higgs boson in association with a photon
in proton-proton collisions at the LHC.
Detailed description of event generation, 
simulation of detector effects, selection of events and the analysis strategy are given in Section \ref{sec:analysis}.
The estimates of the sensitivity that could
be achieved at the LHC for the considered operators are presented in section \ref{sec:results}.
The constraints are obtained using the photon transverse momentum distribution.
Section \ref{sum} summarizes our conclusions.

%
\section{Theoretical framework}\label{sec:framework}
%

In the framework of SM effective field theory, 
the effects of new physics are anticipated to show up
as new interactions among the SM fields. 
These new interactions are suppressed by the inverse powers of  $\Lambda$ which is the characteristic scale of
new physics.  In this approach,  all heavy new degrees of freedom are
integrated out.
Respecting the SM gauge symmetries, Lorentz invariance and lepton and baryon number conservation,  
these effects are parameterized by higher dimension operators with not-known Wilson coefficients.
The dominant contributions to the observable at the LHC are coming from dimension-six operators.
As a result, the Lagrangian of the SM effective theory  can be written as following \cite{r5,r6,r7}:
\begin{eqnarray}
L_{eff} = L_{\rm SM} + \sum_{i}\frac{c_{i}O_{i}}{\Lambda^{2}},
\end{eqnarray}

The dimension-six operators  could be classified and represented in a convenient form
 by choosing a basis of independent dimension-six operators $O_{i}$. 
In this work, our focus is on the operators which are relevant to Higgs+$\gamma$
process according to  the strongly interacting light Higgs
(SILH) basis.  The SILH Lagrangian, $\mathcal{L}_{\rm SILH}$, has the following form \cite{r37,r38}:
\begin{eqnarray}\label{leff}
	\begin{split}
		\mathcal{L}_{\rm SILH} = & \
		\frac{g_s^2\ \bar c_{g}}{m_{W}^2} \Phi^\dag \Phi G_{\mu\nu}^a G_a^{\mu\nu}
		+\frac{g'^2\ \bar c_{\gamma}}{m_{W}^2} \Phi^\dag \Phi B_{\mu\nu} B^{\mu\nu}     
		+ \frac{i g'\ \bar c_{B}}{2 m_{W}^2} \big[\Phi^\dag \overleftrightarrow{D}^\mu \Phi \big] \partial^\nu  B_{\mu \nu}  \\
		& \
		+ \frac{i g\ \bar  c_{W}}{2m_{W}^2} \big[ \Phi^\dag \sigma_{k} \overleftrightarrow{D}^\mu \Phi \big]  D^\nu  W_{\mu \nu}^k    
		+ \frac{ i g\ \bar c_{HW}}{m_{W}^2} \big[D^\mu \Phi^\dag \sigma_{k} D^\nu \Phi\big] W_{\mu \nu}^k  - \frac{\bar c_{6} \lambda}{v^2} \big[\Phi^\dag \Phi \big]^3  \\
		& \
		+ \frac{i g'\ \bar c_{HB}}{m_{W}^2}  \big[D^\mu \Phi^\dag D^\nu \Phi\big] B_{\mu \nu}   
		+ \frac{\bar c_{H}}{2 v^2} \partial^\mu\big[\Phi^\dag \Phi\big] \partial_\mu \big[ \Phi^\dagger \Phi \big]
		+ \frac{\bar c_{T}}{2 v^2} \big[ \Phi^\dag {\overleftrightarrow{D}}^\mu \Phi \big] \big[ \Phi^\dag {\overleftrightarrow{D}}_\mu \Phi \big]    \\
		& \
	    - \bigg[
		\frac{\bar c_{l}}{v^2} y_\ell\ \Phi^\dag \Phi\ \Phi {\bar L}_L e_R
		+\frac{\bar c_{u}}{v^2} y_u \Phi^\dag \Phi\ \Phi^\dag\cdot{\bar Q}_L u_R   
		+ \frac{\bar c_{d}}{v^2} y_d \Phi^\dag \Phi\ \Phi {\bar Q}_L d_R
		+ {\rm h.c.} \bigg]  \,,
	\end{split}
\end{eqnarray}
where $B^{\mu\nu}$,  $W^{\mu \nu}$, $G^{\mu\nu}$ are the 
electroweak and strong field strength tensors and $\Phi$ is a weak doublet which contains the Higgs boson field.
In the above Lagrangian, 
$\Phi^\dag \overleftrightarrow{D}^\mu \Phi = \Phi^\dag (D^{\mu}\Phi) - (D^{\mu}\Phi)^{\dag}\Phi$ and the
 Higgs boson is considered as a CP-even particle.

Very precisely measured  oblique parameters $T$ and $S$ from the electroweak precision measurements
reduce the number of free parameters in the SILH Lagrangian. In particular, the per-mille bounds on $T$ and $S$ parameters 
strongly constrain $\bar c_{T}$ and $\bar c_{B} + \bar c_{W}$ as they are directly related to $T$ and $S$ parameters~\cite{r20,r37,r38,r39}. 
In Ref.\cite{r20}, constraints at $95\%$ confidence level (CL) on the coefficients of dimension six
operators have been obtained using a global fit to the LEP and  the LHC Run I measurements.

This effective Lagrangian is written in gauge basis and after the electroweak symmetry breaking, it can be expressed in the mass 
basis.  In the mass basis, the relevant subset of the Higgs boson CP-conserving anomalous
couplings in unitary gauge includes \cite{r13}:
\begin{eqnarray} \label{ccc}
	\mathcal{L} &=& -\frac{1}{4}g_{h\gamma\gamma}F_{\mu\nu}F^{\mu\nu}h 
	 -\frac{1}{2}g^{(1)}_{h\gamma z}Z_{\mu\nu}F^{\mu\nu}h-g^{(2)}_{h\gamma z}Z_{\nu} \partial_{\mu} F^{\mu\nu}h \nonumber \\
	&-&(\frac{\tilde{y}_{u}}{\sqrt{2}}[\bar{u}P_{R} u] h + \frac{\tilde{y}_{d}}{\sqrt{2}}[\bar{d}P_{R} d] h \nonumber \\
	&+& g^{(\partial)}_{h\gamma u u}[\bar{u}\gamma^{\mu \nu} P_{R} u]F_{\mu\nu}h + g^{(\partial)}_{h\gamma d d}[\bar{d}\gamma^{\mu \nu} P_{R} d]F_{\mu\nu}h + h.c. ), \nonumber \\
\end{eqnarray}
where photon and $Z$-boson field strength tensors denote with $F_{\mu\nu}$ and $Z_{\mu\nu}$, respectively.
Table \ref{tab:coupling} presents the relations between the couplings in Eq.\ref{ccc} and the Wilson
coefficients in SILH Lagrangian (Eq.\ref{leff}). The SM loop induced
contribution to the Higgs boson coupling to two photons ($H\gamma\gamma$)  is denoted by $a_{H}$.

 
 \begin{table}[h!]
 	\caption{\label{tab:coupling} The relations between the anomalous Higgs boson couplings  in the mass and gauge  bases.  
	$a_{H}$ denotes the SM contribution of Higgs boson couplings to two photons at loop level.}
 	\begin{center}
 		\begin{tabular}{|c|c|}
 			\hline
 			Mass basis &  Gauge basis 
 			\\
 			\hline
 			$g_{h\gamma\gamma}$ &   $a_{H} - \frac{8g s^2_{W}}{m_{W}}\bar{c}_{\gamma}$ 
 			\\
 	 		\hline
 			$g^{(1)}_{h\gamma z}$ &     $\frac{g s_{W}}{c_{W}m_{W}}(\bar{c}_{HW}-\bar{c}_{HB}+8 s^2_{W}\bar{c}_{\gamma})$
 			\\
 			\hline
 			$g^{(2)}_{h \gamma z}$ &  $\frac{g s_{W}}{c_{W}m_{W}}(\bar{c}_{HW}-\bar{c}_{HB}-\bar{c}_{B}+\bar{c}_{W})$  
 			\\
 			\hline
 			$\tilde{y}_{u}$ & $y_{u}[1-\frac{1}{2}\bar{c}_{H}+\frac{3}{2}\bar{c}_{u}]$
 			\\
 			\hline
 			$\tilde{y}_{d}$ & $y_{d}[1-\frac{1}{2}\bar{c}_{H}+\frac{3}{2}\bar{c}_{d}]$  
 			\\
 			\hline
 			$g^{(\partial)}_{h \gamma u u }$ &   $\frac{\sqrt{2} g s_{W}}{m^2_{W}}y_{u}[\bar{c}_{uB}+\bar{c}_{uW}]$  
 			\\
 			\hline
 			$g^{(\partial)}_{h \gamma d d }$ & $\frac{\sqrt{2} g s_{W}}{m^2_{W}}y_{d}[\bar{c}_{dB}-\bar{c}_{dW}]$  
 			\\
 			\hline
 		\end{tabular}
 	\end{center}
 \end{table}
 
 \section{Higgs boson production in association with a photon at the LHC}\label{sec:smm}

\begin{figure}[htb]
\begin{center}
\vspace{1cm}
\resizebox{0.6\textwidth}{!}{\includegraphics{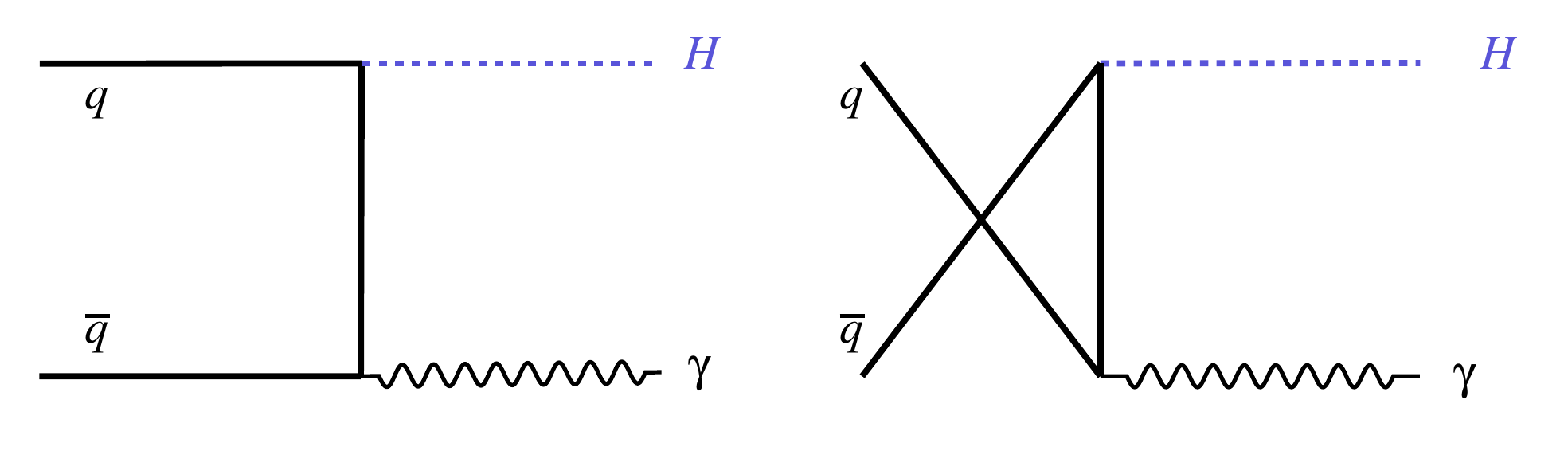}}  
\caption{ Representative Feynman diagrams for Higgs+$\gamma$ production in the SM at leading order.  }\label{feynmanlo}
\end{center}
\end{figure}

 In this section, we discuss the production of a Higgs boson in association with a photon 
 in proton-proton collisions at the LHC. 
 In the SM, at leading order, the H+$\gamma$ production proceeds through
 the quark-antiquark annihilation.  The representative Feynman diagrams are depicted in Fig.\ref{feynmanlo}.

\begin{figure*}[htb]
	\begin{center}
		\vspace{0.50cm}
		\resizebox{0.65\textwidth}{!}{\includegraphics{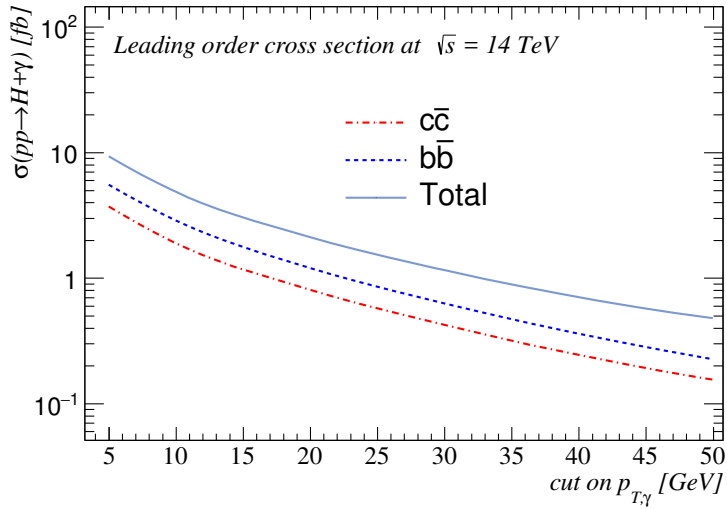}}   
			\caption{ Figure shows the cross section for production of H+$\gamma$ at leading order versus the lower cut 
			 on photon transverse momentum at the LHC with the center-of-mass energy of 14 TeV. The contributions from 
			 $b\bar{b}$, $c\bar{c}$ annihilations are depicted separately.}\label{xs}
	\end{center}
\end{figure*}

The analytical expression for the differential cross section of the partonic process $q\bar{q}\rightarrow H+\gamma$ within the SM framework
 is given by \cite{r340}:
\begin{eqnarray}\label{xsec}
\frac{d\hat{\sigma}}{d\hat{t}} = \frac{\alpha_{em}Q_{q}^{2}\lambda_{q}^{2}}{12}\times\Big[\frac{\hat{t}\hat{u}(1+r_{H}^{2})+8r_{q}^{2}(\hat{t}+\hat{u})^{2}-2r_{q}(4\hat{t}\hat{u}+r_{H}(\hat{t}+\hat{u})^{2})}{(1-4r_{q})\times \hat{t}^{2}\hat{u}^{2}} \Big],
\end{eqnarray}
where
\begin{eqnarray}
\hat{t}=(p_{b}-p_{\gamma})^{2}-m_{b}^{2},~\hat{u}=(p_{\bar{b}}-p_{\gamma})^{2}-m_{b}^{2}, r_{H}=\frac{m_{H}^{2}}{s},~r_{q}=\frac{m_{q}^{2}}{s},
\end{eqnarray}
$s = (p_{b}+p_{\bar{b}})^{2}$, the quark electric charge is denoted by $Q_{q}$, and $\lambda_{q}=\frac{m_{q}}{v}$ is the 
Yukawa coupling of quark $q$.
As it can be seen in Eq.\ref{xsec}, the presence of the electromagnetic coupling constant $\alpha_{em}$ and the Yukawa couplings 
of all light flavor quarks are suppressing factors in the production cross section of Higgs boson in association with a photon.
The main source of tree level contributions come from $c\bar{c}$ and $b\bar{b}$ annihilation because of their larger
Yukawa couplings. However, the contributions of  $c\bar{c}$ and $b\bar{b}$ are suppressed by parton distribution 
function effects with respect to the lighter quarks.
For more illustration, we show the production cross section of Higgs boson in association with a photon  in terms of the lower cut 
 on photon transverse momentum in Fig. \ref{xs}. The contributions from 
$b\bar{b}$, $c\bar{c}$ annihilations are depicted separately.

At  next-to-leading order, the production of H+$\gamma$ through gluon-gluon fusion is forbidden
using the Furry's theorem because of the C parity conservation.
The Higgs boson in association with a photon could be also produced at  next-to-leading order
via quark-antiquark annihilation \cite{r40}.  These processes occur  through box and triangle diagrams in which
$W,Z$, Higgs, and quarks are in the loop. The cross sections of these processes are found to be smaller than the 
leading-order quark-antiquark annihilation so  we neglect them in the present work.
 
\begin{figure}[htb]
\begin{center}
\vspace{1cm}
\resizebox{0.6\textwidth}{!}{\includegraphics{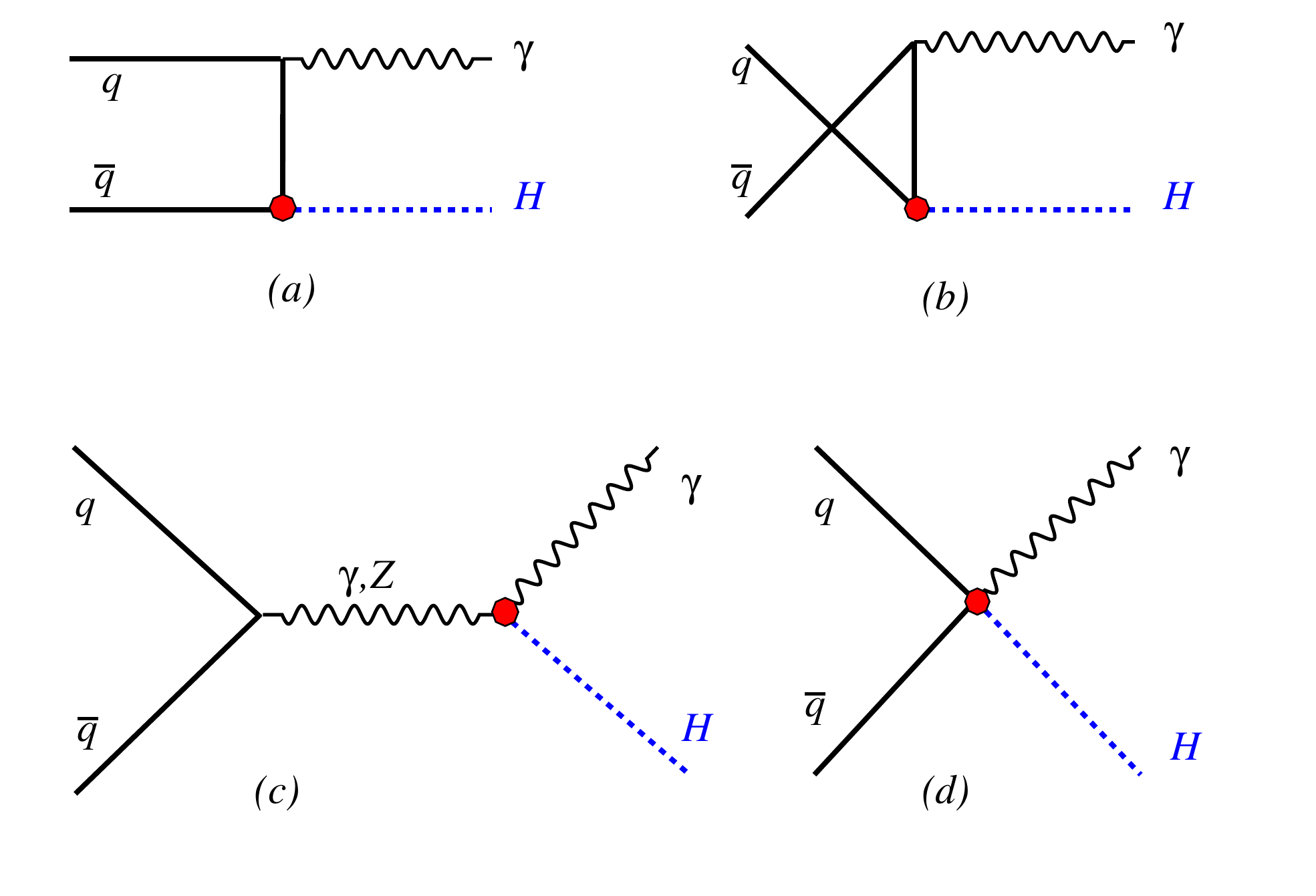}}  
\caption{  Representative Feynman diagrams at tree-level for  production of a Higgs boson in association with a photon at
the LHC in the presence of dimension six operators. }\label{feynmanEFT}
\end{center}
\end{figure}

Within the SM effective field theory, in addition to new diagrams, the SM diagrams are modified as well.
The representative Feynman diagrams for H+$\gamma$ production are depicted
in Fig. \ref{feynmanEFT}. The vertices which receive contributions from the SM effective field theory are shown by
filled circles.  As it can be seen, in addition to the SM tree level diagrams, additional diagrams depicted in the bottom 
of Fig.\ref{feynmanEFT} contribute to the Higgs boson plus a photon at the LHC.

Considering the SM effective Lagrangian, the process $pp\rightarrow H+\gamma$  is sensitive to 
$\bar{c}_{\gamma}$, $\bar{c}_{HW}$,$ \bar{c}_{W}$,$ \bar{c}_{B}$, $\bar{c}_{HB}$,$ \bar{c}_{H}$,$ \bar{c}_{u}$, $\bar{c}_{d}$,
$ \bar{c}_{uW}$,$ \bar{c}_{dW}$,$ \bar{c}_{uB}$,$ \bar{c}_{dB}$
parameters in gauge basis and to  $g_{h\gamma\gamma}$, $g_{h\gamma z}^{(1)}$, 
$g_{h\gamma z}^{(2)}$,  $\tilde{y}_{u}$, $\tilde{y}_{d}$, $g_{h\gamma uu}^{(\partial)}$, $g_{h\gamma dd}^{(\partial)}$ couplings
in the mass basis. As shown in Table \ref{tab:coupling}, the latter four couplings are proportional to the Yukawa couplings of the quarks which are
small for the light quarks.
Therefore, the  sub-processes shown in Fig.\ref{feynmanEFT} (a),(b),(d)  
do not lead to any significant modifications in the H+$\gamma$ production cross section and are not taken into account in this analysis.
As a consequence, we concentrate only on the remaining four parameters: $\bar{c}_{\gamma}$, $\bar{c}_{HW}$,
$\bar{c}_{W}$, and $\bar{c}_{HB}$.

\begin{figure*}[htb]
	\begin{center}
		\vspace{0.50cm}
		\resizebox{0.65\textwidth}{!}{\includegraphics{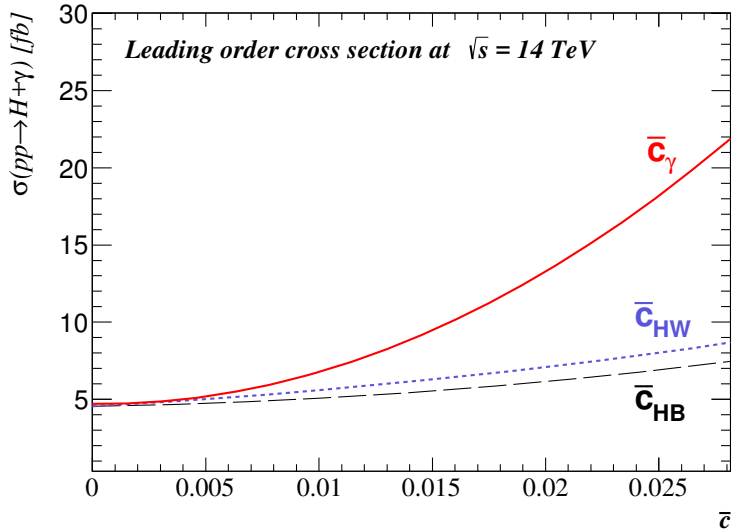}}   
			\caption{ The leading order cross section for production of H+$\gamma$ versus 
			  the coefficients $\bar{c}_{\gamma}$, $\bar{c}_{HB}$, and $\bar{c}_{HW}$ at the LHC with the center-of-mass energy of 14 TeV is shown. 
			  A lower cut of 10 GeV on the photon transverse momentum has been applied.}\label{xss}
	\end{center}
\end{figure*}

Figure \ref{xss} shows the production cross section of H+$\gamma$ in terms of the Wilson coefficients
 $\bar{c}_{\gamma}$, $\bar{c}_{HB}$, and $\bar{c}_{HW}$. To calculate the cross sections, a lower cut of 
 10 GeV has been set on the photon transverse momentum. As it can be seen, there is a significant sensitivity 
 to $\bar{c}_{\gamma}$ while $\bar{c}_{HB}$ and $\bar{c}_{HW}$ have smaller effects on the production rate.
It should be mentioned that $\bar{c}_{W}$  and $\bar{c}_{B}$ are  found to have no
big effect in the cross section.

In this work,  the effects of the dimension six operators on H+$\gamma$ production are calculated at the LHC
using { \tt MadGraph5-aMC@NLO}~\cite{r444,r445}.
The effective SM Lagrangian presented in Eq.~\ref{leff} has been implemented in {\tt FeynRule}
program \cite{r446} and then the UFO model \cite{ufo} is inserted into { \tt MadGraph5-aMC@NLO}. 
The  details could be found in Refs.~\cite{r13}.
We will discuss the details of simulations and determination of the  95\% CL
constraints on the coefficients of dimension six operators in the next section.

%
\section{Simulation details and analysis strategy}\label{sec:analysis}

In this section, we present the simulation details and the analysis strategy 
for exploring the effective SM in Higgs boson production associated with a photon.
Due to large branching fraction, we consider the Higgs boson decay into a pair of bottom quarks.
Consequently, the final state consists of  one energetic photon together with two jets originating from the hadronization of  b-quarks.

The main sources of background processes which are taken into account in this work are W$^{\pm}\gamma$+jets,
Z$\gamma$+jets, $\gamma$+jets, top+$\gamma$+jets, $t\bar{t}+\gamma$.
{\tt MadGraph5-aMC@NLO} event generator is used to generate the 
effective SM signal and background processes 
~\cite{r444,r445}.  We generate events at the center-of-mass energy of 14 TeV at the LHC  using
the NNPDF2.3 \cite{nnpdf} set of parton distribution functions. 
In the event generation process, the SM input parameters are taken as \cite{Agashe:2014kda}:
$m_W = 80.385 \, {\rm GeV}, m_H = 125.0 \, {\rm GeV}, \, m_t = 173.34 \, {\rm GeV} $ and $m_Z = 91.187 \, {\rm GeV}$.
The events are passed through {\tt PYTHIA 8}~\cite{Sjostrand:2007gs,Sjostrand:2003wg} for
showering of partons, hadronization, and unstable particles decays. {\tt Delphes 3.3.2}
~\cite{deFavereau:2013fsa,Mertens:2015kba} allows to simulate a CMS-like detector effects.

Jets are reconstructed using  anti-$k_{t}$ algorithm~\cite{Cacciari:2008gp} with a radius parameter of R = 0.5. 
This is performed through the {\tt FastJet}  package~\cite{Cacciari:2011ma}.
The efficiency of b-tagging and the misidentification rates are taken to be dependent on the transverse momentum of jets
according to the following formulas \cite{btag}:
\begin{eqnarray}
\epsilon_{b}(p_{T}) = 0.85\times \tanh(0.0025\times p_{T})&\times& \big(\frac{25.0}{1+0.063\times p_{T}} \big),\nonumber \\
\text{mis-id. rate of c-jets} = 0.25 \times \tanh(0.018\times p_{T})&\times& \big(\frac{1.0}{1+0.0013\times p_{T}} \big),\\ \nonumber
\text{mis-id. rate of light-jets} = 0.01+0.000038 &\times& p_{T}.
\end{eqnarray}
The b-tagging efficiency for a jet with $p_{T}=50$ GeV is around $64\%$, while the c-jet misidentification rate is 
$17\%$, and light flavor jets misidentification rate is $1.1\%$.

To select signal events, we require that photon has
a transverse momentum $p_{T}$ and pseudorapidity $\eta$
satisfying $p_{T} > 40$ GeV and $|\eta| < 2.5$. Any event 
containing a charged lepton with $p_{T} > 10$ GeV and $|\eta| < 2.5$ or missing transverse energy above 30 GeV is vetoed.
Events are selected by demanding the presence of two b-jets with $p_{T} > 20$ GeV and $|\eta| < 2.5$.
To ensure all selected objects are well-isolated, the angular separation 
$\Delta R(\gamma, {\rm b-jets}) = \sqrt{(\Delta \phi)^2 + (\Delta \eta)^2} > 0.4$ and $\Delta R({\rm b_{1}}, {\rm b_{2}})  > 0.4$. 
The above described selection are denoted as preselection cuts.

\begin{figure*}[htb]
	\begin{center}
		\vspace{0.50cm}
		\resizebox{0.46\textwidth}{!}{\includegraphics{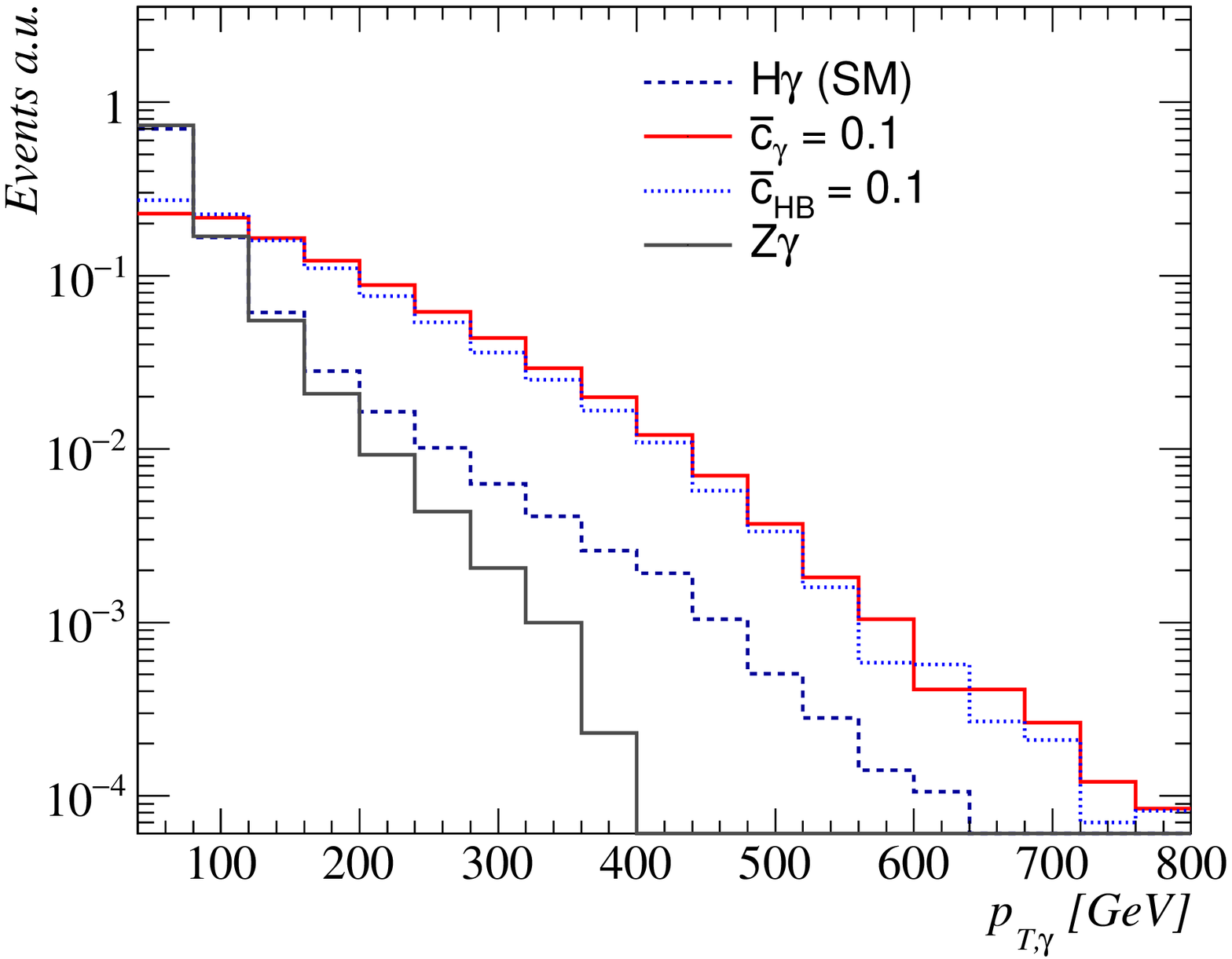}}  
		\resizebox{0.432\textwidth}{!}{\includegraphics{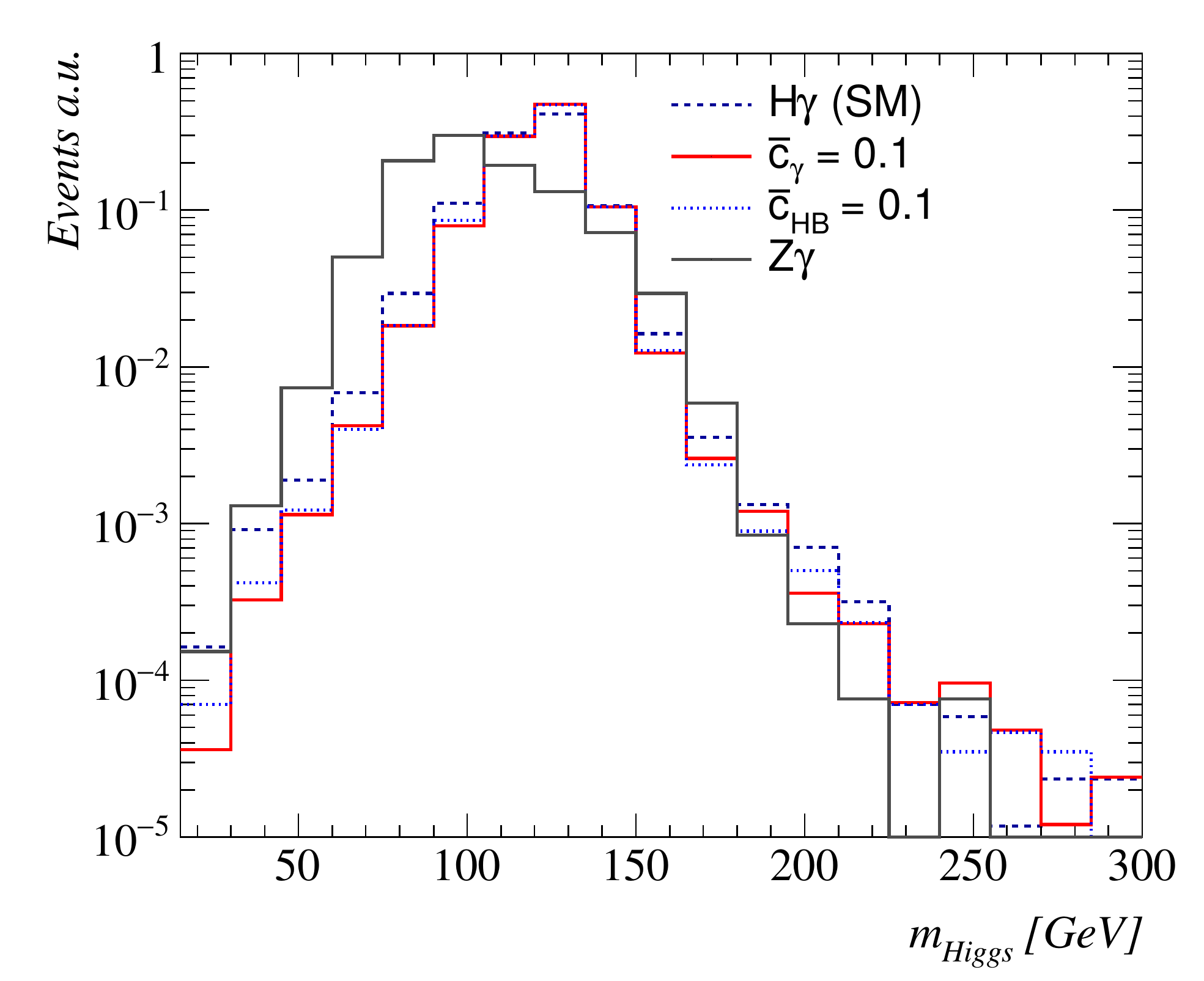}}   
		\resizebox{0.440\textwidth}{!}{\includegraphics{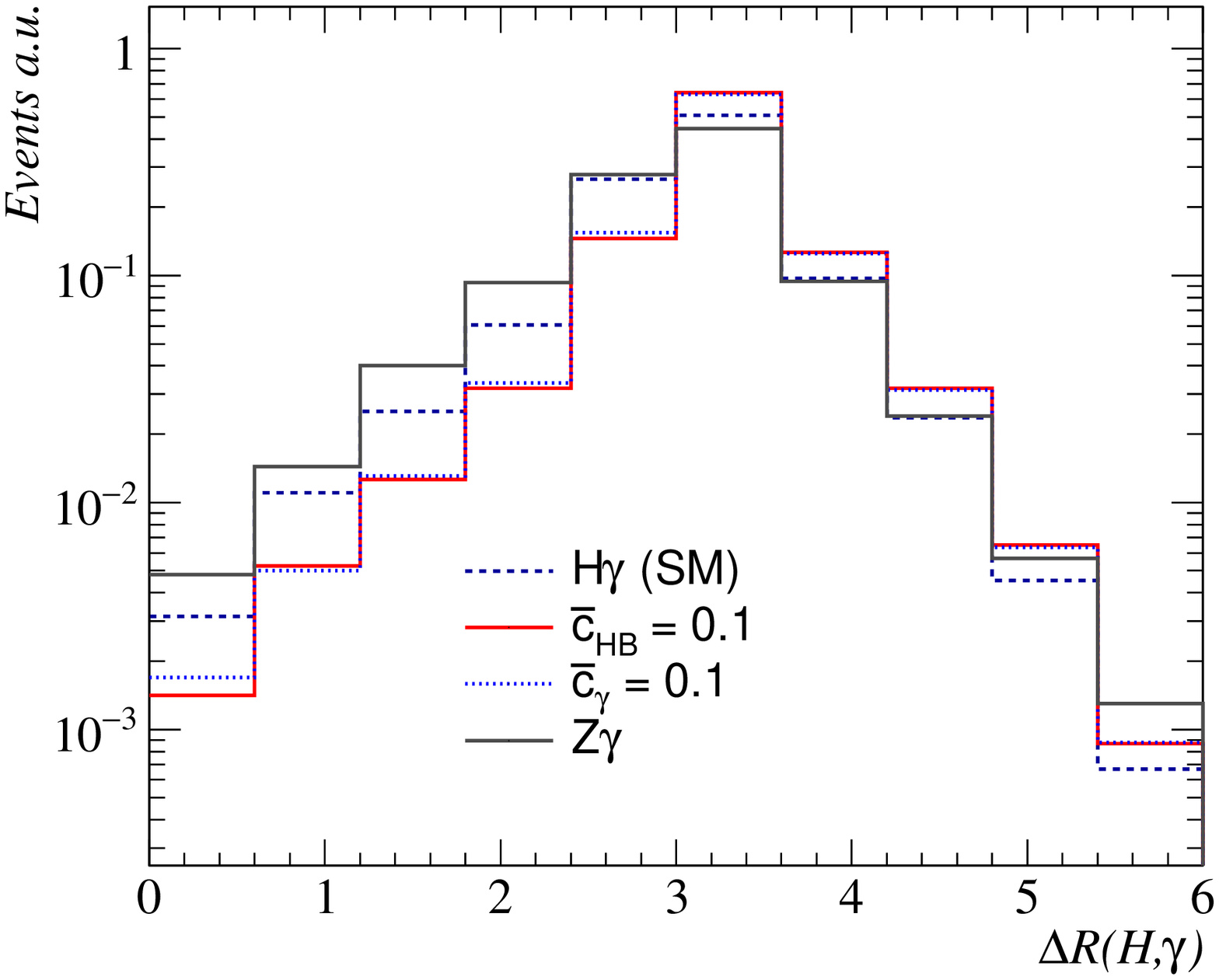}}   
		\resizebox{0.450\textwidth}{!}{\includegraphics{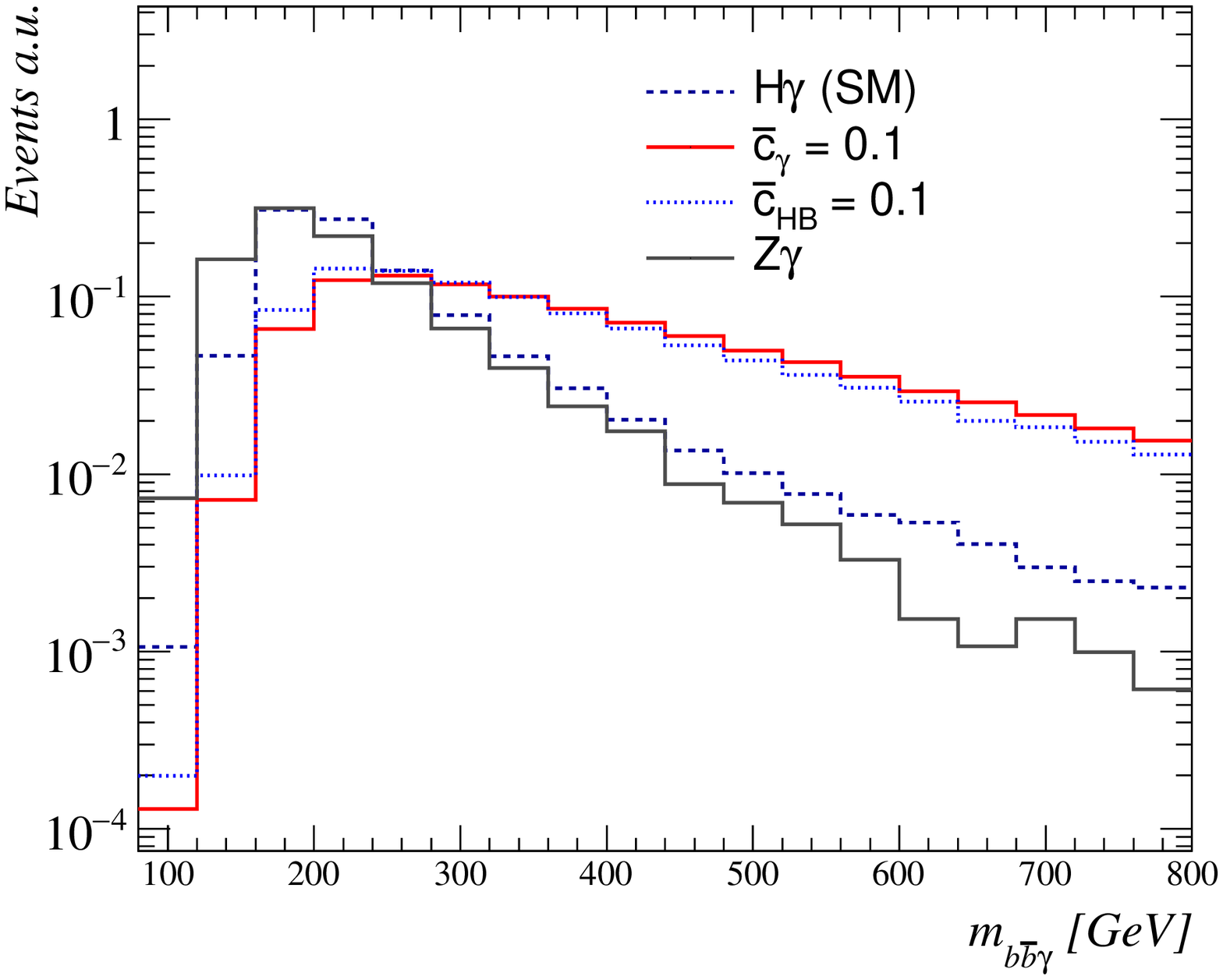}}   
		\caption{ Plots show the transverse momentum 
of the photon (upper left), the mass of reconstructed Higgs boson (upper right), the $\Delta R(H,\gamma)$ (lower left)
, and the invariant mass of the H+$\gamma$ system (lower right) for
the  SM H+$\gamma$, effective SM with
 $\bar{c}_{\gamma} = 0.1$ and $\bar{c}_{HB} = 0.1$ and $Z\gamma$ background process.}\label{plots}
	\end{center}
\end{figure*}

In Figure \ref{plots}, we present the transverse momentum 
of the photon (upper left),  mass of the reconstructed Higgs boson from $b\bar{b}$ pair (upper right), the $\Delta R(H,\gamma)$ (lower left)
, and the invariant mass of the H+$\gamma$ system (lower right) for
the irreducible SM background process Z$\gamma$, SM H+$\gamma$, and for effective SM with
 $\bar{c}_{\gamma} = 0.1$ and $\bar{c}_{HB} = 0.1$.
 All these distributions are depicted after applying the so called preselection cuts described previously. 

From the photon transverse momentum spectrum and invariant mass of the H+$\gamma$ system, one can see that 
due to the presence of derivatives in the induced couplings of the effective SM, 
significant enhancements in the tail of photon $p_{T}$ and $m_{H\gamma}$ appear.  
This is a specific feature of new effective interactions  which allows us to differentiate between new physics and SM background processes.


For further reduction of the  SM background contributions additional cuts are applied.
A window cut on the reconstructed Higgs boson mass leads to suppress backgrounds 
without a Higgs boson in the final state. Thus, it is required that $90 < m_{b \bar{b}} < 160$ GeV.
The angular separation between the reconstructed Higgs boson and photon candidate is required to 
be greater than 2.4 which is useful to suppress the $\gamma$+jets background process. 
A cut on the invariant mass of H+$\gamma$ reduces the contribution of all background processes.  

Cross sections of signal for the cases of $\bar{c}_{\gamma} = 0.1$,  $\bar{c}_{HB} = 0.1$
and background processes after applying each set of cuts are shown in Table~\ref{Table:Cut-Table}. The cross sections
are given in the unit of fb. 
Since the background contributions overwhelm the signal in the lower values of cut on photon transverse momentum,
we concentrate on  the region where  the signal-to-background ratio is large enough to determine the exclusion limits.

\begin{table}[ht]
	\begin{center} 
		\caption{ Expected cross sections in unit of fb after different combinations of cuts for signal and SM background processes. The signal cross sections
	 are corresponding to particular values of $\bar{c}_{\gamma} = 0.1$, $\bar{c}_{HB} = 0.1$. More details of the cuts are given in the text. }
		\begin{tabular}{c|cc|c c c c}
			$\sqrt s = 14$ TeV      &  \multicolumn{2}{c|}{~Signal }    & \multicolumn{3}{c}{\,\,\,\,\,~~~~~~~~~ Background processes }    \\   \hline
			Cuts    & $\bar{c}_\gamma$     &  $\bar{c}_{HB}$ &     $\gamma$+jets &  $tj\gamma+t\bar{t}\gamma$ & $W \gamma+Z \gamma$ & SM (H$\gamma$) \\ \hline  \hline
			One photon and  lepton veto   &  $126.3$  &  $20.99$     &   $4.443 \times 10^6$   &  $164.92$     & $10286.4$   &  $0.450$ \\		
			Only 2\,{\rm b-jets}, $\Delta R_{i,j}  > 0.4; i,j=\gamma,b$       &  $19.77$   & $3.41$       &  $13026.2$   &  $0.0312$  & $119.57$    &  $0.080$ \\ \hline \hline
			$90 <  m_{b \bar{b}} < 160$ GeV and $\Delta R (H, \gamma) > 2.4$ &  $17.88$   &  $3.06$    &   $6397.6$   & $0.0165$     &  $58.41$    &  $0.067$ \\ 
	      	$m_{b \bar{b} \gamma } > 250$ and $p_T^{\gamma} > 400$  GeV         &  $0.51$   &  $0.076$ &    $0.0$   & $0.0$   &  $0.0$  &    $0.0003$ \\ \hline  \hline
		\end{tabular}
	\label{Table:Cut-Table}
		\end{center}
\end{table}

At the end of this section, a discussion is given on the potential contribution
of background coming from multijet production where jets are misidentified as photons. 
Such a signature could come up when in the jet fragmentation,  neutral pions appear and decay 
into two photons with a large boost. As a result, the two photon showers will overlap in the 
electromagnetic calorimeter and will not be separately distinguishable and they appear as a single photon 
in the detector.
The ability for rejecting jets faking photons is crucial to suppress the QCD multijet background
which has several orders of magnitude larger cross section with respect to the other background processes.
After the kinematic requirements (except those related to photon) in our analysis as well as the requirement of the presence of only
two b-jets, the cross section of multijet background reduces to around $10^{3}$ pb. 
The jet fake photon probability varies with the fake photon transverse momentum and is at the order of $10^{-5,-4}$
for the high $p_{T}$ fake photons ($p_{T} > 200$ GeV). Consequently, the requirement of a photon with large 
transverse momentum decreases the contribution of this background to a negligible level. 
We neglect this background in this analysis nevertheless a dedicated and more realistic detector simulation
must be performed in order to estimate possible contribution of this background.

%
\section{  Results and interpretation}\label{sec:results}
%

\begin{figure*}[htb]
	\begin{center}
		\vspace{0.450cm}
		\resizebox{0.6\textwidth}{!}{\includegraphics{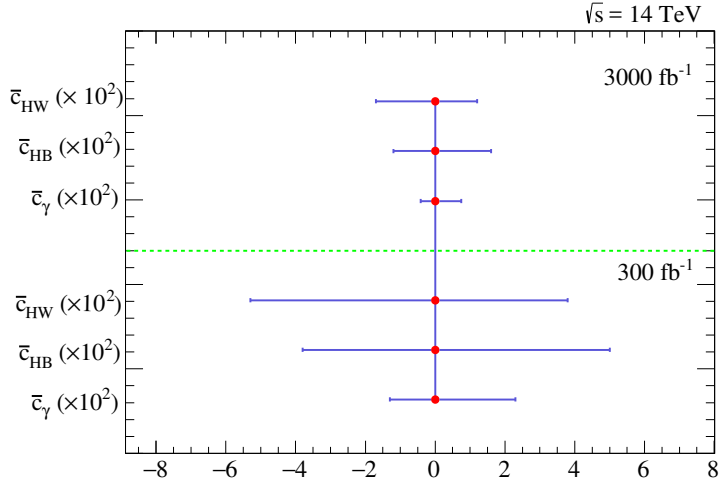}}   
		\caption{ The $95\%$ CL ranges obtained in a fit to the dimension-six operator coefficients individually (setting the others to zero). 
             The results are presented for the integrated luminosities of 300 fb$^{-1}$ and 3 ab$^{-1}$. }\label{results_plot}
	\end{center}
\end{figure*}

In this section, we present the potential sensitivities of the Higgs boson production in association with a photon to 
the  dimension-six operators. 
The main coefficients affecting the Higgs+$\gamma$ production that are our
interest in this analysis are $\bar{c}_{\gamma}$, $\bar{c}_{HW}$, and $\bar{c}_{HB}$.
Constraints derivation is based on the fact that dimension-six operators generate a higher tail at the 
large values of photon transverse momentum.  As a result, we make use of the tail of photon transverse momentum, where 
the contribution of the SM background processes are negligible, to obtain the limits on the relevant dimension-six operators. 
As the number of signal events increases quickly with the $\bar{c}_{\gamma}$, $\bar{c}_{HW}$, and $\bar{c}_{HB}$
in the tail of $p_{T,\gamma}$ and the background contribution is significantly suppressed, a $\chi^{2}$ fit is performed in this 
region to get the $95\%$ CL ranges.  To obtain the predicted limits, only one coefficient is considered in the fit.
The predicted bounds at $95\%$ CL for the LHC at $\sqrt{s}=14$ TeV with an integrated
luminosity of 300 fb$^{-1}$  and 3000 fb$^{-1}$ are presented in Table \ref{tab:results1}. To obtain the results, 
only statistical uncertainties are considered and the effects of systematic and theoretical sources 
are neglected. 
The results are also depicted in Fig.\ref{results_plot}  for the integrated luminosities 
of 300 and 3000 fb$^{-1}$.
As it can be seen, the most sensitivity is to $\bar{c}_{\gamma}$ which is expected to be probed
down to $10^{-3}$ while $\bar{c}_{HB}$ and $\bar{c}_{HW}$ are potentially explored 
at the order of $10^{-2}$ at the $95\%$ CL. 

At this point it should be mentioned that based on Fig.\ref{plots} (lower right), the information from 
the tail of the invariant mass distribution of H+$\gamma$ 
could also be used to extract the constraints on the coefficients. 
As it can be seen, while the expectation of the SM steeply drops for the invariant mass values larger than
around 300 GeV,  the new physics effects shows a tail extending up to around the TeV scale. 
However, in a realistic analysis,
more systematic uncertainties with respect to the photon transverse momentum distribution enters
in the analysis using $m_{H\gamma}$ differential distribution.

\begin{table}[htbp]
\begin{center}
\begin{tabular}{c|c|c}
    	\hline
	 Coefficient & $\mathcal{L}=300$ fb$^{-1}$ & $\mathcal{L}=3000$ fb$^{-1}$ \\ 
	\hline
	$\bar{c}_{\gamma}$    &   $[\, -0.013, \, 0.023 \, ]$  &  $[\, -0.0042, \, 0.0075 \, ]$ \\
	$\bar{c}_{HB}$       &      $[\, -0.038, \, 0.050 \, ]$  &  $[\, -0.012, \, 0.016 \, ]$ \\
	$\bar{c}_{HW}$       &      $[\, -0.053, \, 0.038 \, ]$  & $[\, -0.017, \, 0.012 \, ]$ \\
	\hline 
\end{tabular}
	\caption{ Predicted constraints at 95\% CL on dimension-six operator coefficients for the LHC with 14 TeV with 
		the integrated luminosities of 300 fb$^{-1}$ and 3000  fb$^{-1}$. 
		At a time one of the couplings is 
		 considered in the analysis. \label{tab:results1} }
		\end{center}
\end{table}

It is informative to compare the results obtained in this study with the 
expected bounds at high luminosity LHC with other channels.
In Ref.\cite{r50}, considering the production modes $pp\rightarrow H+j$, 
$pp\rightarrow H+2j$, $pp\rightarrow H$, $pp\rightarrow W+H$,
$pp\rightarrow Z+H$, and $pp\rightarrow t\bar{t}+H$, constraints 
are set on the dimension-six operator coefficients.  The following
bounds have been obtained on $\bar{c}_{\gamma}$, $\bar{c}_{HW}$, $\bar{c}_{HB}$ at 95$\%$ CL  
at the center-of-mass energy of 14 TeV
with 3000 fb$^{-1}$ based on the expected signal strength:
\begin{eqnarray}
-0.027 < \bar{c}_{HW} <  0.028, ~  -0.026 < \bar{c}_{HB} < 0.027,~  -0.00029 < \bar{c}_{\gamma} < 0.00027,
\end{eqnarray}
The limits on $\bar{c}_{HW}$ and $\bar{c}_{HB}$ are the same order of the ones could be achieved from this study.  However, they could be
 improved using a shape analysis on the Higgs transverse momentum as:
\begin{eqnarray}
-0.004 < \bar{c}_{HW} <  0.004, ~  -0.004 < \bar{c}_{HB} < 0.004,~  -0.00016 < \bar{c}_{\gamma} < 0.00013,
\end{eqnarray}
It should be mentioned that  two improvements to the limits on dimension-six coefficients would be 
the combination of H+$\gamma$ channel with the other channels discussed in Ref.\cite{r50} as well as
considering the complete next-to-leading order predictions for the process
of H+$\gamma$ with the dimension-six couplings.

\begin{figure*}[htb]
	\begin{center}
		\vspace{0.50cm}
		\resizebox{0.5\textwidth}{!}{\includegraphics{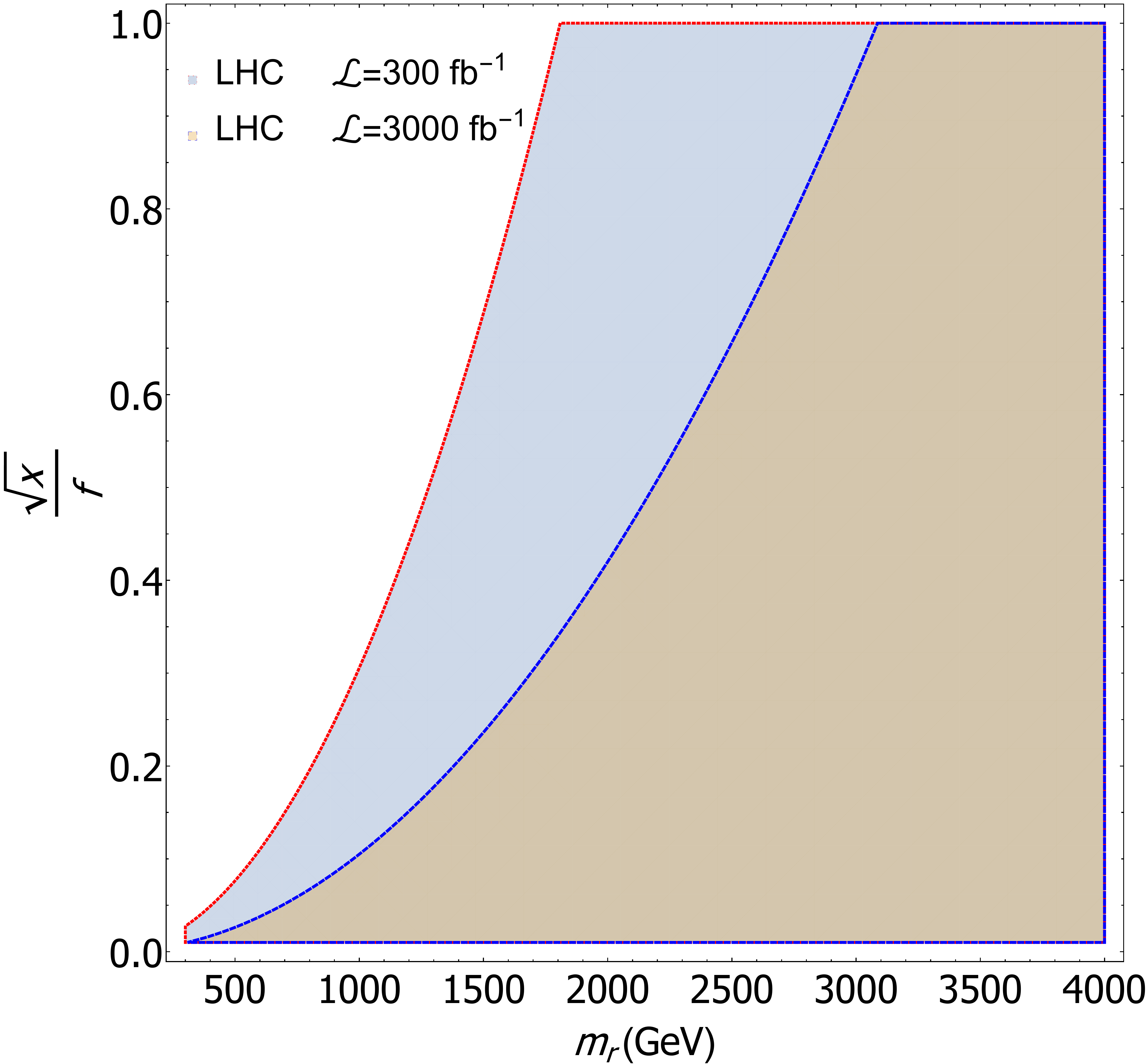}}   
		\caption{ The $95\%$ CL ranges obtained on the dilaton  mass and the dilaton field coupling obtained 
from the exclusion range of $\bar{c}_{\gamma}$.  The parameter  $x$ is defined as $ \frac{g}{g'}\times \sqrt{b_{1}\alpha_{1}}$.}\label{reach}
	\end{center}
\end{figure*}

As it is well-known, the effective field theory configuration provides the possibility to constrain UV models parameters.
In Ref.\cite{uv1}, the matching of dimension-six operators with few UV models are studied.  The discussed UV models 
are corresponding to extension of the Higgs sector of the SM with adding a scalar doublet (two-Higgs doublet models), or with 
a radion or a dilaton or with including an extra scalar singlet.  
By matching the dimension-six operators to the UV theories, one could see that 
an operator can be generated in these models at tree level, at one loop or at higher orders.
For example, the exchange of a radion or a dilaton scalar
generates $\bar{c}_{\gamma}$, $\bar{c}_{HW}$, $\bar{c}_{HB}$, $\bar{c}_{H}$, $\bar{c}_{6}$, 
$\bar{c}_{T}$, $\bar{c}_{W}$, $\bar{c}_{B}$, and $\bar{c}_{g}$ at tree level.
 A radion could arise from 
 the excitation of  graviton in extra dimensions and a dilaton come from spontaneous breaking of scale invariance. 
 Detailed description of the models could be found in \cite{uv1,uv2,r53}.
 The coefficients of dimension-six operators are related to the radion/dilaton model parameters according to 
 the following:
 \begin{eqnarray}
\bar{c}_{\gamma} = -\frac{b_{1}\alpha_{1}g^{2}}{4\pi g'^{2}}\frac{m_{H}^{2}v^{2}}{f^{2}m_{r}^{2}},~
\bar{c}_{HB} = -\frac{b_{1}\alpha_{1}}{4}\frac{m_{H}^{2}v^{2}}{f^{2}m_{r}^{2}},~
\bar{c}_{HW} = -\frac{b_{2}\alpha_{2}}{4 }\frac{m_{H}^{2}v^{2}}{f^{2}m_{r}^{2}},~
\bar{c}_{H} = \frac{8m_{H}^{2}v^{2}}{f^{2}m_{r}^{2}}.
\end{eqnarray}
where  $b_{i}$ are the coefficients of $\beta$ function which for dilatons violate the scale invariance anomalously and for radions are the
localized fields contribution near the IR brane \cite{uv1,uv2,r53}.  The couplings constants corresponding to the SM gauge groups 
$SU(3)\times SU(2) \times U(1)$ are denoted by $\alpha_{1,2,3}$. $m_{r}$ and $f$ are the dilaton/radion mass and compactification scale (spontaneous
symmetry breaking scale). 
Fig.\ref{reach} shows the $95\%$ CL exclusion region of the dilaton  mass and the dilaton field coupling obtained 
from the exclusion range of $\bar{c}_{\gamma}$. In this plot, the parameter $x = \frac{g}{g'}\times \sqrt{b_{1}\alpha_{1}}$. The regions are
depicted for the LHC with 300 fb$^{-1}$ and 3 ab$^{-1}$. We see that LHC 
would be able to extend the reach for the mass of radion to TeV scale (depending on the couplings and  the value of $f$).

\section{Summary and outlook}
\label{sum}

According to the recent measurements of the LHC experiments, Higgs boson 
properties are found to be consistent with the SM predictions within the uncertainties and so far no considerable 
sign of new physics has been found.  As a result, in order to search for the possible effects of new physics
one can focus on the SM effective theory which relies on dimension-six operators whose contributions are
suppressed by the second power of new physics scale.  
In the present work, we have studied the impact of dimension-six operators 
of the SM effective field theory related to the Higgs boson production in association with a photon at the LHC. 
In particular, the search is performed in the SILH basis in which a
complete set of dimension-six operators with minimum assumptions on the Wilson coefficients is considered.
We have restricted ourselves to the set of CP-even operators which affect the Higgs+$\gamma$ production at tree level
and concentrate only the Higgs boson decays into $b\bar{b}$ pairs.
The dimension-six operators which contain derivative interactions modify the kinematics of Higgs+$\gamma$
 production. Specially,  the tails of photon transverse momentum and the invariant mass of the
 Higgs+$\gamma$ ($m_{H\gamma}$) receive significant contributions from the new couplings. 
 Considering a fast detector simulation with {\tt Delphes} and the main sources of background processes,
 stringent bounds on the coefficients of dimension-six operators have been obtained using the
information of the tail of photon transverse momentum. It is found that the H+$\gamma$ production
 allows us to probe $\bar{c}_{\gamma}$ and  $\bar{c}_{HW} (\bar{c}_{HB})$
down to $10^{-3}$ and $10^{-2}$, respectively, with an integrated luminosity of 3000 fb$^{-1}$.

Matching between the Wilson coefficients and UV models, we have the opportunity to probe the
parameters of the UV models. In particular, we have constrained the radion/dilaton parameters and 
found out the mass of dilaton up to the scale of few TeV is accessible.

There are possibilities for improvement of our results listed in the following. 
The first way of improvement is to use the complete next-to-leading order predictions for the process
of H+$\gamma$ with the dimension-six couplings \cite{r5555}.  Second  possibility is the  incorporation  of  the H+$\gamma$+jet
 process  \cite{r54,r55} into  our analysis which would increase the sensitivity.  The third way of increasing the sensitivity is the inclusion of
 other decay modes of the Higgs boson 
 such as $WW$, $ZZ$, etc. which allows to have more statistics and possibly leads to more stringent bounds. At the end, 
 including  the Higgs+$\gamma$ process in a global fit with other production processes at the LHC would affect the
 exclusion ranges and may improve the sensitivities.

%
\section*{Acknowledgments}
%

Hamzeh Khanpour is thankful to School of Particles and Accelerators, Institute for Research in Fundamental Sciences (IPM), and 
University of Science and Technology of Mazandaran for financial support of this project.
Sara Khatibi would like to thank Iran National Science Foundation (INSF) for
the financial support.


%

\begin{thebibliography}{}
%

\bibitem{atlash} 
  G.~Aad {\it et al.} [ATLAS Collaboration],
  Phys.\ Lett.\ B {\bf 716}, 1 (2012)
  doi:10.1016/j.physletb.2012.08.020
  [arXiv:1207.7214 [hep-ex]].


\bibitem{cmsh} 
  S.~Chatrchyan {\it et al.} [CMS Collaboration],
  Phys.\ Lett.\ B {\bf 716}, 30 (2012)
  doi:10.1016/j.physletb.2012.08.021
  [arXiv:1207.7235 [hep-ex]].

\bibitem{susy} 
  H.~P.~Nilles,
  Phys.\ Rept.\  {\bf 110}, 1 (1984).
  doi:10.1016/0370-1573(84)90008-5


	
\bibitem{r1} 
  S.~Chatrchyan {\it et al.} [CMS Collaboration],
  JHEP {\bf 1306}, 081 (2013)
  doi:10.1007/JHEP06(2013)081
  [arXiv:1303.4571 [hep-ex]].

\bibitem{r2} 
 G.~Aad {\it et al.} [ATLAS Collaboration],
  Phys.\ Lett.\ B {\bf 726}, 88 (2013)
  Erratum: [Phys.\ Lett.\ B {\bf 734}, 406 (2014)]
  doi:10.1016/j.physletb.2014.05.011, 10.1016/j.physletb.2013.08.010
  [arXiv:1307.1427 [hep-ex]].

\bibitem{r220} 
 A.~Djouadi and G.~Moreau,
  Eur.\ Phys.\ J.\ C {\bf 73}, no. 9, 2512 (2013)
  doi:10.1140/epjc/s10052-013-2512-9
  [arXiv:1303.6591 [hep-ph]].

\bibitem{r221} 
  D.~Carmi, A.~Falkowski, E.~Kuflik, T.~Volansky and J.~Zupan,
  JHEP {\bf 1210}, 196 (2012)
  doi:10.1007/JHEP10(2012)196
  [arXiv:1207.1718 [hep-ph]].



\bibitem{r3} 
 K.~G.~Wilson,
  Rev.\ Mod.\ Phys.\  {\bf 55}, 583 (1983).
  doi:10.1103/RevModPhys.55.583


\bibitem{r4} 
 T.~Appelquist and J.~Carazzone,
  Phys.\ Rev.\ D {\bf 11}, 2856 (1975).
  doi:10.1103/PhysRevD.11.2856

\bibitem{r5} 
 W.~Buchmuller and D.~Wyler,
  Nucl.\ Phys.\ B {\bf 268}, 621 (1986).
  doi:10.1016/0550-3213(86)90262-2

\bibitem{r6} 
B.~Grzadkowski, M.~Iskrzynski, M.~Misiak and J.~Rosiek,
  JHEP {\bf 1010}, 085 (2010)
  doi:10.1007/JHEP10(2010)085
  [arXiv:1008.4884 [hep-ph]].


\bibitem{r7} 
 K.~Hagiwara, S.~Ishihara, R.~Szalapski and D.~Zeppenfeld,
  Phys.\ Rev.\ D {\bf 48}, 2182 (1993).
  doi:10.1103/PhysRevD.48.2182

\bibitem{r8} 
 C.~N.~Leung, S.~T.~Love and S.~Rao,
  Z.\ Phys.\ C {\bf 31}, 433 (1986).
  doi:10.1007/BF01588041


\bibitem{r9} 
 M.~B.~Einhorn and J.~Wudka,
  Nucl.\ Phys.\ B {\bf 876}, 556 (2013)
  doi:10.1016/j.nuclphysb.2013.08.023
  [arXiv:1307.0478 [hep-ph]].

\bibitem{r10} 
 S.~Willenbrock and C.~Zhang,
  Ann.\ Rev.\ Nucl.\ Part.\ Sci.\  {\bf 64}, 83 (2014)
  doi:10.1146/annurev-nucl-102313-025623
  [arXiv:1401.0470 [hep-ph]].


\bibitem{r11} 
  R.~Contino, M.~Ghezzi, C.~Grojean, M.~Muhlleitner and M.~Spira,
  JHEP {\bf 1307}, 035 (2013)
  doi:10.1007/JHEP07(2013)035
  [arXiv:1303.3876 [hep-ph]].

\bibitem{r12} 
R.~S.~Gupta, A.~Pomarol and F.~Riva,
  Phys.\ Rev.\ D {\bf 91}, no. 3, 035001 (2015)
  doi:10.1103/PhysRevD.91.035001
  [arXiv:1405.0181 [hep-ph]].


\bibitem{r13} 
 A.~Alloul, B.~Fuks and V.~Sanz,
  JHEP {\bf 1404}, 110 (2014)
  doi:10.1007/JHEP04(2014)110
  [arXiv:1310.5150 [hep-ph]].

\bibitem{r14} 
 J.~R.~Espinosa, C.~Grojean, M.~Muhlleitner and M.~Trott,
  JHEP {\bf 1212}, 045 (2012)
  doi:10.1007/JHEP12(2012)045
  [arXiv:1207.1717 [hep-ph]].

\bibitem{r15} 
  T.~Corbett, O.~J.~P.~Eboli, J.~Gonzalez-Fraile and M.~C.~Gonzalez-Garcia,
  Phys.\ Rev.\ D {\bf 87}, 015022 (2013)
  doi:10.1103/PhysRevD.87.015022
  [arXiv:1211.4580 [hep-ph]].


\bibitem{r16} 
 M.~E.~Peskin,
  arXiv:1207.2516 [hep-ph].

\bibitem{r17} 
B.~Dumont, S.~Fichet and G.~von Gersdorff,
  JHEP {\bf 1307}, 065 (2013)
  doi:10.1007/JHEP07(2013)065
  [arXiv:1304.3369 [hep-ph]].


\bibitem{r18} 
  D.~Lopez-Val, T.~Plehn and M.~Rauch,
  JHEP {\bf 1310}, 134 (2013)
  doi:10.1007/JHEP10(2013)134
  [arXiv:1308.1979 [hep-ph]].


\bibitem{r19} 
 C.~Englert, A.~Freitas, M.~M.~Muhlleitner, T.~Plehn, M.~Rauch, M.~Spira and K.~Walz,
  J.\ Phys.\ G {\bf 41}, 113001 (2014)
  doi:10.1088/0954-3899/41/11/113001
  [arXiv:1403.7191 [hep-ph]].


\bibitem{r20} 
  J.~Ellis, V.~Sanz and T.~You,
  JHEP {\bf 1503}, 157 (2015)
  doi:10.1007/JHEP03(2015)157
  [arXiv:1410.7703 [hep-ph]].


\bibitem{r21} 
 A.~Falkowski and F.~Riva,
  JHEP {\bf 1502}, 039 (2015)
  doi:10.1007/JHEP02(2015)039
  [arXiv:1411.0669 [hep-ph]].


\bibitem{r22} 
 T.~Corbett, O.~J.~P.~Eboli, D.~Goncalves, J.~Gonzalez-Fraile, T.~Plehn and M.~Rauch,
  JHEP {\bf 1508}, 156 (2015)
  doi:10.1007/JHEP08(2015)156
  [arXiv:1505.05516 [hep-ph]].

\bibitem{r23} 
  G.~Buchalla, O.~Cata, A.~Celis and C.~Krause,
  Eur.\ Phys.\ J.\ C {\bf 76}, no. 5, 233 (2016)
  doi:10.1140/epjc/s10052-016-4086-9
  [arXiv:1511.00988 [hep-ph]].


\bibitem{r24} 
  T.~Plehn and M.~Rauch,
  Europhys.\ Lett.\  {\bf 100}, 11002 (2012)
  doi:10.1209/0295-5075/100/11002
  [arXiv:1207.6108 [hep-ph]].



\bibitem{r25} 
 T.~Corbett, O.~J.~P.~Eboli, J.~Gonzalez-Fraile and M.~C.~Gonzalez-Garcia,
  Phys.\ Rev.\ D {\bf 86}, 075013 (2012)
  doi:10.1103/PhysRevD.86.075013
  [arXiv:1207.1344 [hep-ph]].


\bibitem{r26} 
 J.~Ellis, V.~Sanz and T.~You,
  JHEP {\bf 1407}, 036 (2014)
  doi:10.1007/JHEP07(2014)036
  [arXiv:1404.3667 [hep-ph]].


\bibitem{r27} 
F.~Ferreira, B.~Fuks, V.~Sanz and D.~Sengupta,
  arXiv:1612.01808 [hep-ph].


\bibitem{r28} 
 A.~Azatov, R.~Contino and J.~Galloway,
  JHEP {\bf 1204}, 127 (2012)
  Erratum: [JHEP {\bf 1304}, 140 (2013)]
  doi:10.1007/JHEP04(2012)127, 10.1007/JHEP04(2013)140
  [arXiv:1202.3415 [hep-ph]].


\bibitem{r29} 
  L.~Berthier and M.~Trott,
  JHEP {\bf 1602}, 069 (2016)
  doi:10.1007/JHEP02(2016)069
  [arXiv:1508.05060 [hep-ph]].


\bibitem{r30} 
  A.~Efrati, A.~Falkowski and Y.~Soreq,
  JHEP {\bf 1507}, 018 (2015)
  doi:10.1007/JHEP07(2015)018
  [arXiv:1503.07872 [hep-ph]].


\bibitem{r31} 
H.~Bélusca-Maïto, A.~Falkowski, D.~Fontes, J.~C.~Romão and J.~P.~Silva,
  arXiv:1611.01112 [hep-ph].


\bibitem{r32} 
 S.~Bar-Shalom, A.~Soni and J.~Wudka,
  Phys.\ Rev.\ D {\bf 92}, no. 1, 015018 (2015)
  doi:10.1103/PhysRevD.92.015018
  [arXiv:1405.2924 [hep-ph]].


\bibitem{r33} 
 J.~Cohen, S.~Bar-Shalom and G.~Eilam,
  Phys.\ Rev.\ D {\bf 94}, no. 3, 035030 (2016)
  doi:10.1103/PhysRevD.94.035030
  [arXiv:1602.01698 [hep-ph]].


\bibitem{r330}
  C.~Englert, Y.~Soreq and M.~Spannowsky,
  JHEP {\bf 1505}, 145 (2015)
  doi:10.1007/JHEP05(2015)145
  [arXiv:1410.5440 [hep-ph]].


\bibitem{r331}
 C.~Englert and M.~Spannowsky,
  Phys.\ Lett.\ B {\bf 740}, 8 (2015)
  doi:10.1016/j.physletb.2014.11.035
  [arXiv:1408.5147 [hep-ph]].

\bibitem{r88}
 M.~J.~Dolan, C.~Englert, N.~Greiner, K.~Nordstrom and M.~Spannowsky,
  Eur.\ Phys.\ J.\ C {\bf 75}, no. 8, 387 (2015)
  doi:10.1140/epjc/s10052-015-3622-3
  [arXiv:1506.08008 [hep-ph]].

\bibitem{r89}
  M.~Buschmann, C.~Englert, D.~Goncalves, T.~Plehn and M.~Spannowsky,
  Phys.\ Rev.\ D {\bf 90}, no. 1, 013010 (2014)
  doi:10.1103/PhysRevD.90.013010
  [arXiv:1405.7651 [hep-ph]].

\bibitem{r999}
 S.~Taheri Monfared, S.~Fayazbakhsh and M.~Mohammadi Najafabadi,
  Phys.\ Lett.\ B {\bf 762}, 301 (2016)
  doi:10.1016/j.physletb.2016.09.055
  [arXiv:1610.02883 [hep-ph]].


\bibitem{r1000} 
  S.~Banerjee, T.~Mandal, B.~Mellado and B.~Mukhopadhyaya,
  JHEP {\bf 1509}, 057 (2015)
  doi:10.1007/JHEP09(2015)057
  [arXiv:1505.00226 [hep-ph]].

\bibitem{r1001} 
 S.~Kumar, P.~Poulose and S.~Sahoo,
  Phys.\ Rev.\ D {\bf 91}, no. 7, 073016 (2015)
  doi:10.1103/PhysRevD.91.073016
  [arXiv:1501.03283 [hep-ph]].


\bibitem{r1002} 
F.~Maltoni, E.~Vryonidou and C.~Zhang,
  JHEP {\bf 1610}, 123 (2016)
  doi:10.1007/JHEP10(2016)123
  [arXiv:1607.05330 [hep-ph]].


\bibitem{r1003} 
 R.~M.~Godbole, D.~J.~Miller, K.~A.~Mohan and C.~D.~White,
  JHEP {\bf 1504}, 103 (2015)
  doi:10.1007/JHEP04(2015)103
  [arXiv:1409.5449 [hep-ph]].


\bibitem{r1004} 
 E.~Boos, V.~Bunichev, M.~Dubinin and Y.~Kurihara,
  Phys.\ Rev.\ D {\bf 89}, 035001 (2014)
  doi:10.1103/PhysRevD.89.035001
  [arXiv:1309.5410 [hep-ph]].

\bibitem{r1005} 
M.~B.~Einhorn and J.~Wudka,
  Nucl.\ Phys.\ B {\bf 877}, 792 (2013)
  doi:10.1016/j.nuclphysb.2013.11.004
  [arXiv:1308.2255 [hep-ph]].


\bibitem{r1006} 
 G.~Isidori and M.~Trott,
  JHEP {\bf 1402}, 082 (2014)
  doi:10.1007/JHEP02(2014)082
  [arXiv:1307.4051 [hep-ph]].

\bibitem{r1007}
 S.~Khatibi and M.~Mohammadi Najafabadi,
  Phys.\ Rev.\ D {\bf 90}, no. 7, 074014 (2014)
  doi:10.1103/PhysRevD.90.074014
  [arXiv:1409.6553 [hep-ph]].


\bibitem{r1008}
  A.~Kobakhidze, N.~Liu, L.~Wu and J.~Yue,
  Phys.\ Rev.\ D {\bf 95}, no. 1, 015016 (2017)
  doi:10.1103/PhysRevD.95.015016
  [arXiv:1610.06676 [hep-ph]].


\bibitem{r1009}
  H.~Khanpour and M.~Mohammadi Najafabadi,
  Phys.\ Rev.\ D {\bf 95}, no. 5, 055026 (2017)
  doi:10.1103/PhysRevD.95.055026
  [arXiv:1702.00951 [hep-ph]].

\bibitem{r34} 
 D.~de Florian {\it et al.} [LHC Higgs Cross Section Working Group],
  arXiv:1610.07922 [hep-ph].


\bibitem{r340}
 E.~Gabrielli, B.~Mele and J.~Rathsman,
  Phys.\ Rev.\ D {\bf 77}, 015007 (2008)
  doi:10.1103/PhysRevD.77.015007
  [arXiv:0707.0797 [hep-ph]].



\bibitem{r35} 
 G.~Buchalla, O.~Cata and C.~Krause,
  Nucl.\ Phys.\ B {\bf 894}, 602 (2015)
  doi:10.1016/j.nuclphysb.2015.03.024
  [arXiv:1412.6356 [hep-ph]].



\bibitem{r36} 
 G.~Buchalla, O.~Cata, A.~Celis and C.~Krause,
  Phys.\ Lett.\ B {\bf 750}, 298 (2015)
  doi:10.1016/j.physletb.2015.09.027
  [arXiv:1504.01707 [hep-ph]].


\bibitem{r37} 
 R.~Contino, M.~Ghezzi, C.~Grojean, M.~Muhlleitner and M.~Spira,
  JHEP {\bf 1307}, 035 (2013)
  doi:10.1007/JHEP07(2013)035
  [arXiv:1303.3876 [hep-ph]].


\bibitem{r38} 
 G.~F.~Giudice, C.~Grojean, A.~Pomarol and R.~Rattazzi,
  JHEP {\bf 0706}, 045 (2007)
  doi:10.1088/1126-6708/2007/06/045
  [hep-ph/0703164].


\bibitem{r39} 
R.~Barbieri, A.~Pomarol, R.~Rattazzi and A.~Strumia,
Nucl.\ Phys.\ B {\bf 703}, 127 (2004)
doi:10.1016/j.nuclphysb.2004.10.014
[hep-ph/0405040].



\bibitem{r40} 
  A.~Abbasabadi, D.~Bowser-Chao, D.~A.~Dicus and W.~W.~Repko,
  Phys.\ Rev.\ D {\bf 58}, 057301 (1998)
  doi:10.1103/PhysRevD.58.057301
  [hep-ph/9706335].


\bibitem{r444}
 J.~Alwall {\it et al.},
  JHEP {\bf 1407}, 079 (2014)
  doi:10.1007/JHEP07(2014)079
  [arXiv:1405.0301 [hep-ph]].

\bibitem{r445}
 J.~Alwall, M.~Herquet, F.~Maltoni, O.~Mattelaer and T.~Stelzer,
  JHEP {\bf 1106}, 128 (2011)
  doi:10.1007/JHEP06(2011)128
  [arXiv:1106.0522 [hep-ph]].


\bibitem{r446}
  A.~Alloul, N.~D.~Christensen, C.~Degrande, C.~Duhr and B.~Fuks,
  Comput.\ Phys.\ Commun.\  {\bf 185}, 2250 (2014)
  doi:10.1016/j.cpc.2014.04.012
  [arXiv:1310.1921 [hep-ph]].


\bibitem{ufo} 
  C.~Degrande, C.~Duhr, B.~Fuks, D.~Grellscheid, O.~Mattelaer and T.~Reiter,
  Comput.\ Phys.\ Commun.\  {\bf 183}, 1201 (2012)
  doi:10.1016/j.cpc.2012.01.022
  [arXiv:1108.2040 [hep-ph]].


\bibitem{r41} 
P.~Artoisenet {\it et al.},
JHEP {\bf 1311}, 043 (2013)
doi:10.1007/JHEP11(2013)043
[arXiv:1306.6464 [hep-ph]].
		

\bibitem{Sjostrand:2007gs} 
T.~Sjostrand, S.~Mrenna and P.~Z.~Skands,
Comput.\ Phys.\ Commun.\  {\bf 178}, 852 (2008)
doi:10.1016/j.cpc.2008.01.036
[arXiv:0710.3820 [hep-ph]].
	
	
	
\bibitem{Sjostrand:2003wg} 
T.~Sjostrand, L.~Lonnblad, S.~Mrenna and P.~Z.~Skands,
hep-ph/0308153.

	

\bibitem{deFavereau:2013fsa} 
J.~de Favereau {\it et al.} [DELPHES 3 Collaboration],
JHEP {\bf 1402}, 057 (2014)
doi:10.1007/JHEP02(2014)057
[arXiv:1307.6346 [hep-ex]].
	
	
	
	
\bibitem{Mertens:2015kba} 
A.~Mertens,
J.\ Phys.\ Conf.\ Ser.\  {\bf 608}, no. 1, 012045 (2015).
doi:10.1088/1742-6596/608/1/012045
	
	

\bibitem{nnpdf}
  R.~D.~Ball {\it et al.},
  Nucl.\ Phys.\ B {\bf 867}, 244 (2013)
  doi:10.1016/j.nuclphysb.2012.10.003
  [arXiv:1207.1303 [hep-ph]].



\bibitem{Agashe:2014kda} 
K.~A.~Olive {\it et al.} [Particle Data Group Collaboration],
Chin.\ Phys.\ C {\bf 38}, 090001 (2014).
doi:10.1088/1674-1137/38/9/090001
	
	
	
\bibitem{Cacciari:2008gp} 
M.~Cacciari, G.~P.~Salam and G.~Soyez,
JHEP {\bf 0804}, 063 (2008)
doi:10.1088/1126-6708/2008/04/063
[arXiv:0802.1189 [hep-ph]].

	
	
\bibitem{Cacciari:2011ma} 
M.~Cacciari, G.~P.~Salam and G.~Soyez,
Eur.\ Phys.\ J.\ C {\bf 72}, 1896 (2012)
doi:10.1140/epjc/s10052-012-1896-2
[arXiv:1111.6097 [hep-ph]].
	
\bibitem{btag}
 S.~Chatrchyan {\it et al.} [CMS Collaboration],
  JINST {\bf 8}, P04013 (2013)
  doi:10.1088/1748-0221/8/04/P04013
  [arXiv:1211.4462 [hep-ex]].


\bibitem{r50}
 C.~Englert, R.~Kogler, H.~Schulz and M.~Spannowsky,
  Eur.\ Phys.\ J.\ C {\bf 76}, no. 7, 393 (2016)
  doi:10.1140/epjc/s10052-016-4227-1
  [arXiv:1511.05170 [hep-ph]].

	
\bibitem{uv1}
 M.~Gorbahn, J.~M.~No and V.~Sanz,
  JHEP {\bf 1510}, 036 (2015)
  doi:10.1007/JHEP10(2015)036
  [arXiv:1502.07352 [hep-ph]].


\bibitem{uv2}
  C.~Csaki, M.~L.~Graesser and G.~D.~Kribs,
  Phys.\ Rev.\ D {\bf 63}, 065002 (2001)
  doi:10.1103/PhysRevD.63.065002
  [hep-th/0008151].


\bibitem{r53}
  W.~D.~Goldberger, B.~Grinstein and W.~Skiba,
  Phys.\ Rev.\ Lett.\  {\bf 100}, 111802 (2008)
  doi:10.1103/PhysRevLett.100.111802
  [arXiv:0708.1463 [hep-ph]].

\bibitem{r5555}
 C.~Degrande, B.~Fuks, K.~Mawatari, K.~Mimasu and V.~Sanz,
  arXiv:1609.04833 [hep-ph].


\bibitem{r54}
 P.~Agrawal and A.~Shivaji,
  Phys.\ Lett.\ B {\bf 741}, 111 (2015)
  doi:10.1016/j.physletb.2014.12.021
  [arXiv:1409.8059 [hep-ph]].


\bibitem{r55}
  E.~Gabrielli, B.~Mele, F.~Piccinini and R.~Pittau,
  JHEP {\bf 1607}, 003 (2016)
  doi:10.1007/JHEP07(2016)003
  [arXiv:1601.03635 [hep-ph]].


		
	
\end{thebibliography}
%


%

\end{document}